\documentclass[journal,10pt]{IEEEtran}
\IEEEoverridecommandlockouts
\usepackage[utf8]{inputenc}
\usepackage[T1]{fontenc}
\usepackage{amsmath,amssymb,amsfonts,times}
\usepackage{graphicx}
\usepackage{textcomp}
\usepackage{pgfplots}
\usepackage{scalefnt}
\newcommand{\norm}[1]{\left\lVert#1\right\rVert}

\def\BibTeX{{\rm B\kern-.05em{\sc i\kern-.025em b}\kern-.08em
    T\kern-.1667em\lower.7ex\hbox{E}\kern-.125emX}}
\begin{document}

\title{Study of Max-Link Relay Selection with Buffers for Multi-Way Cooperative Multi-Antenna Systems \\
}

\author{F. L. Duarte
            and R. C. de Lamare,~\IEEEmembership{Senior Member,~IEEE}
\thanks{F. L. Duarte is with the Centre for Telecommunications Studies (CETUC), Pontifical Catholic University of Rio de Janeiro, Brazil, and the Military Institute of Engineering, IME, Rio de Janeiro, RJ, Brazil. e-mail: flaviold@cetuc.puc-rio.br}
\thanks{R. C. de Lamare is with the Centre for Telecommunications Studies (CETUC), Pontifical Catholic University of Rio de Janeiro, Brazil, and the Department of Eletronic Engineering, University of York, United Kingdon. e-mail: delamare@cetuc.puc-rio.br}}

\maketitle
\linespread{0.9}
\begin{abstract}

In this paper, we present a  relay-selection strategy for multi-way
cooperative multi-antenna systems that are aided by a central
processor node, where a cluster formed by two users is selected to
simultaneously transmit to each other with the help of relays. In
particular, we present a novel multi-way relay selection strategy
based on the selection of the best link, exploiting the use of
buffers and physical-layer network coding, that is called Multi-Way
Buffer-Aided Max-Link  (MW-Max-Link). We compare the proposed
MW-Max-Link to existing techniques in terms of bit error rate,
pairwise error probability, sum rate and computational complexity.
Simulations are then employed to evaluate the performance of
the proposed and existing techniques.  \\
\end{abstract}


\section{INTRODUCTION}

\IEEEPARstart{T}{he} Multi-Way Relay Channel \cite{f80} includes a
full data exchange model, in which each user receives messages of
all other users, and the pairwise data exchange model, which
consists of multiple two-way relay channels over which two users
($U_1$ and $U_2$) exchange messages with the help of a common
intermediate relay $R$. In order to adapt to modern requirements,
relaying schemes with high spectrum efficiency have recently
attracted considerable attention \cite{f55,tds_cl,tds}.

An important two-way protocol category is called Multiple-Access
Broadcast-Channel (MABC). In MABC decode-and-forward (DF) protocols,
as in TW-Max-Min \cite{f40}, transmission is organized in a prefixed
schedule with two consecutive time slots. In the first time slot (MA
phase), a selected relay receives and decodes the data
simultaneously transmitted from two source nodes and physical layer
network coding (PLNC) may be employed on the decoded data. In the
second time slot (BC phase), the same relay forwards the decoded
data to the two source nodes, which become destinations. Since all
the channels are reciprocal (restricted to Time Division
Multiplexing - TDM) and fixed during the two phases of the MABC
protocol, the TW-Max-Min  protocol \cite{f40} achieves a maximum
diversity gain. On the other hand, by considering non reciprocal
channels, the performance of relaying schemes may be improved by
using a buffer-aided relaying protocol, where the relay may
accumulate packets in a buffer\cite{f14}, before transmitting to the
destination nodes, as in the one-way Max-Link protocol, which
selects in each time slot the more powerful channel among all the
available source-relay (SR) and relay-destination (RD) channels
(i.e., among $2N$ channels) \cite{f9}. For independent and
identically distributed (i.i.d.) channels, Max-Link achieves a
diversity gain of $2N$, where $N$ is the number of relays. Prior
work has not considered multi-way protocols for multi-antenna
systems or the use of  a multi-way Max-Link (in which each pair of
users has a particular buffer in the relays) or a central processor
node.

In this work, we propose a multi-way Max-Link protocol for
buffer-aided cooperative multi-antenna systems (MW-Max-Link) in non
reciprocal channels. The proposed MW-Max-Link protocol \cite{bamlrs}
selects the best channels among $Z$ pairs of users and achieves a
diversity gain of $2NZ$. We also extend the MMD criterion \cite{f41}
to multi-way systems for selection of relays in the proposed scheme
and the existing TW-Max-Min (here adapted for multi-antenna systems)
and carry out pairwise error probability (PEP), sum rate and
computational complexity analyses.

\section{System Description}

\begin{figure}[!h]
\centering
\includegraphics[scale=0.44]{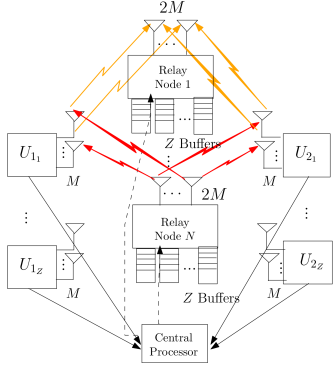}
\caption{System model of a buffer-aided multi-way relay system}
\label{fig:model}
\end{figure}

We consider a multi-antenna multi-way MABC relay scheme with $Z$
pairs of users and $N$ half-duplex DF relays, $R_1$,...,$R_N$
operating in a spatial multiplexing mode. The users are equipped
with $M$ antennas and each relay with $2M$ antennas. A total of $Z$
buffers are accessed by the selected relays for storing or
extracting (each pair of users has a particular buffer established
on demand in the relays), as shown in Fig.\ref{fig:model}. In the MA
phase, a relay $R_g$ will be selected to receive simultaneously $M$
packets from a selected cluster (pair of users $U_1$ and $U_2$) and
decode the data. Then, PLNC is employed on the decoded data and the
resulting packets are stored in their particular buffers. In the BC
phase, a relay $R_{f}$ will be selected to transmit $M$ packets from
the particular buffer to the selected cluster. We remark that
distributed space-time coding can also be examined in the described
framework \cite{armo}.

\subsection{Assumptions}

In each time slot, the total energy transmitted from each user to
the relay selected for reception  or from the relay selected for
transmission to the selected cluster is the same and equal to $E$.
The channel coefficients are drawn from mutually independent zero
mean complex Gaussian random variables. The transmission is
performed in data packets and the channels are constant for the
duration of one packet and vary independently from one packet to the
following. The order of the data packets is inserted in the preamble
of each packet, so the original order is restored at the destination
nodes. Pilot symbols for training and estimation of channel state
information (CSI), and signaling for network coordination are also
inserted in the preamble of the packet \cite{smce,segce}. A central
processor node is responsible for deciding whether a cluster or the
relay should transmit in a given time slot $i$, through a feedback
channel. This can be ensured by an appropriate signalling that
provides global CSI at the central processor node \cite{f9}.
Furthermore, we assume that each relay only has information about
its $U_1R$ and $U_2R$ channels. The use of a unique central
processor node reduces its complexity,  since a single central node
is responsible for deciding which node will transmit (rather than
all destination nodes being responsible together).

\subsection{System Model}

At the MA phase of multi-way MABC DF systems, the received signal
from the selected cluster $U$ (formed by  $U_1$ and $U_2$) to the
selected relay $R_g$  is formed by an $2M \times 1$ vector
$\mathbf{y}_{u,r_g }[i]$ given by
\begin{eqnarray}
   \mathbf{y}_{u,r_g}[i]=\sqrt{\frac{E}{M}} \mathbf{H}_{u,r_g}\mathbf{x}[i]+\mathbf{n}_{r_g}[i],
    \label{eq:2}
\end{eqnarray}
\noindent where $\mathbf{x}[i]$ represents
the vector formed by $M$ symbols transmitted
by $U_1$  and $U_2$ ($\mathbf{x_1}[i]$ and $\mathbf{x_2}[i]$),
$\mathbf{H}_{u,r_g}$ is the $2M \times  2M$ matrix of $U_1R_g$ and $U_2R_g$ channels and
$\mathbf{ n}_{r_g}$ represents the zero mean additive white
complex Gaussian noise (AWGN) at the relay selected for
reception.

Assuming synchronization, we employ the Maximum Likelihood (ML)
receiver at the selected relay for reception:
    \begin{eqnarray}
    \hat{\mathbf{x}}[i]= \arg \min_{\mathbf{x'}[i]} \left(\norm{\mathbf{y}_{u,r_g}[i]- \sqrt{\frac{E}{M}} \mathbf{H}_{u,r_g}\mathbf{x'}[i]}^2\right),
    \label{eq:4}
    \end{eqnarray}
where $\mathbf{x'}[i]$ represents each of the $K^{2M}$  possible
transmitted symbols vector $\mathbf{x}[i]$ ($K$ is the number of
constellation symbols). The ML receiver computes an estimate of the
vector of symbols transmitted by the users $\hat{\mathbf{x}}[i]$.
Considering BPSK ($K= 2$), unit power symbols and $M = 1$, the
estimated symbol vector $ \hat{\mathbf{x}}[i]$ may be $[-1~-1]^T$,
$[-1~+1]^T$, $[+1~-1]^T$ or $[+1~+1]^T$.

By employing PLNC (XOR), it is not necessary to store the $2M$
packets transmitted by the selected cluster, but only the resulting
$M$ packets (XOR outputs) with the information: "the bit transmitted
by $U_1$ is different (or not) from the corresponding bit
transmitted by $U_2$". Then, we employ the XOR:
$\mathbf{v}_{[i]}=\mathbf{\hat{x}_1}[i] \oplus
\mathbf{\hat{x}_2}[i]$ and store the resulting packets in the
buffer.

At the BC phase, the signal transmitted from the selected relay
$R_{f}$ and received at the selected cluster ($U_1$ and $U_2$) is
formed by an $M \times 1$ vector $\mathbf{y}_{r_f,u_{1(2)}}[i]$
given by
    \begin{eqnarray}
    \mathbf{y}_{r_f,u_{1(2)}}[i]=\sqrt{\frac{E}{M}}  \mathbf{H}_{r_f,u_{1(2)}}\mathbf{v}[i]+\mathbf{n}_{u_{1(2)}}[i],
    \label{eq:3}
    \end{eqnarray}
\noindent where $\mathbf{H}_{r_f,u_{1(2)}}$ is the $M \times M$
matrix of  $R_{f}U_{1(2)}$ channels and $\mathbf{n}_{u_{1(2)}}[i]$
represents the AWGN at $U_1$ or $U_2$. At the selected cluster, we
also employ the ML receiver which yields
    \begin{eqnarray}
    \hat{\mathbf{v}}_{1(2)}[i]= \arg \min_{\mathbf{v'}[i]} \left(\norm{\mathbf{y}_{r_f,u_{1(2)}}[i]- \sqrt{ \frac{E}{M}} \mathbf{H}_{r_f,u_{1(2)}}\mathbf{v'}[i]}^2\right)
    \label{eq:6}
    \end{eqnarray}
where $\mathbf{v'}[i]$ represents each possible vector formed by $M$
symbols. Then, at $U_1$ we compute the vector of symbols transmitted
by $U_2$ by employing PLNC (XOR): $\mathbf{\hat{x}_2}[i]$=
$\mathbf{x}_1[i] \oplus \hat{\mathbf{v}}_1[i]$. The same reasoning
is applied at $U_2$ to compute the vector of symbols transmitted by
$U_1$: $\mathbf{\hat{x}_1}[i]$= $\mathbf{x}_2[i] \oplus
\hat{\mathbf{v}}_2[i]$. Considering imperfect CSI when applying the
ML receiver \cite{mldls}, the estimated channel matrix
$\mathbf{\hat{H}}$ is assumed instead of $\mathbf{H}$ in
(\ref{eq:4}) and (\ref{eq:6}). Other suboptimal receivers such as
linear, successive interference cancellation and decision feedback
\cite{windpassinger,spa,mfsic,mbdf} can also be considered.

\section{Proposed MW-Max-Link  Relay Selection Scheme}

The proposed  MW-Max-Link scheme is modelled by the system shown in
Fig. \ref{fig:model}. This proposed scheme operates in two possible
modes in each time slot: MA or  BC. It is not necessary that a
number of the buffer elements be filled with packets before the
system starts its normal operation for this scheme to work properly
and may be empty. Despite that, in this work, we consider that half
of the buffer elements are filled in an initialization phase
\cite{f41}, before the scheme is used. The following subsections
explain how MW-Max-Link works.

\subsection{Relay selection metric}

 In the first step, for each cluster formed by $U_1$ and $U_2$, we compute the metric $\mathcal{B}_{UR_n}$
associated with the user-relay (UR) channels of each relay $R_n$, for the MA mode:
\begin{eqnarray}
  \mathcal{B}_{UR_n}=  \norm{\sqrt{ \frac{E}{M}} \mathbf{H}_{u,r_n}\mathbf{x}_i - \sqrt{ \frac{E}{M}} \mathbf{H}_{u,r_n}\mathbf{x}_j}^2,
  \label{eq:7}
\end{eqnarray}
where $n \in \{1, ...,N\}$, $\mathbf{x}_i$  and $\mathbf{x}_j$
represent each possible vector formed by $2M$ symbols and "$i$" is different from "$j$". This metric is computed for each of the $C_2^{K^{2M}}$
(combination of $K^{2M}$  in $2$) possibilities.

 In the second step, we store the smallest metric ($\mathcal{B}_{min UR_n}$), for being
critical, and thus each relay
will have a minimum distance associated with its UR channels:
\begin{eqnarray}
  \mathcal{B}_{\min UR_n} = \min{(\mathcal{B}_{UR_n})}
\end{eqnarray}

Then, in the third step, we perform ordering on $\mathcal{B}_{\min UR_n}$ and store the largest of these distances:
\begin{eqnarray}
  \mathcal{B}_{z_{\max \min UR}} = \max(\mathcal{B}_{\min UR_n})
\end{eqnarray}
where $z \in \{1, ...,Z\}$. After finding $\mathcal{B}_{z_{\max \min UR}}$  for each cluster, we perform ordering and store the largest of these distances:
\begin{eqnarray}
  \mathcal{B}_{\max \min UR} = \max(\mathcal{B}_{z_{\max \min UR}})
\end{eqnarray}

Then, we select the cluster and the relay that is associated with this distance to receive simultaneously $M$ packets from the selected cluster.

In the fourth step, for each cluster, we compute the metric $\mathcal{B}_{R_{n}U_1}$
associated with the $RU_1$ channels of each relay $R_n$, for the BC mode:
\begin{eqnarray}
    \mathcal{B}_{R_{n}U_1}=  \norm{\sqrt{ \frac{E}{M}} \mathbf{H}_{r_{n},u_1}\mathbf{x}_i - \sqrt{\frac{ E}{M}} \mathbf{H}_{r_{n},u_1}\mathbf{x}_j}^2,
    \label{eq:8a}
\end{eqnarray}
where $n$ $\in \{1, ...,N\}$, $\mathbf{x}_i$  and $\mathbf{x}_j$
represent each possible vector formed by $M$ symbols and "$i$" is different from "$j$". This metric is computed for each of the $C_2^{K^M}$  possibilities. In the fifth step, we find the minimum distance for each relay $R_n$:
\begin{eqnarray}
 \mathcal{B}_{\min R_{n}U_1}=\min{(\mathcal{B}_{R_{n}U_1})},
 \label{eq:9a}
\end{eqnarray}

In the sixth step,  we apply the same reasoning of (\ref{eq:8a}) and (\ref{eq:9a}),  to compute the metrics $\mathcal{B}_{R_{n}U_2}$ and $\mathcal{B}_{\min R_{n}U_2}$. In the seventh step, we compare the distances $\mathcal{B}_{\min R_{n}U_1}$ and $\mathcal{B}_{\min R_nU_2}$  and store the smallest one:
\begin{eqnarray}
\mathcal{B}_{\min R_{n}U} = \min(\mathcal{B}_{\min R_{n}U_1},\mathcal{B}_{\min R_{n}U_2})
\end{eqnarray}

In the eighth step, after finding $\mathcal{B}_{\min R_{n}U}$ for each relay $R_n$, we perform ordering and store the largest of these distances:
\begin{eqnarray}
\mathcal{B}_{z_{\max \min RU}}=\max(\mathcal{B}_{\min R_{n}U})
\end{eqnarray}
where $z \in \{1, ...,Z\}$. After finding $\mathcal{B}_{z_{\max \min RU}}$ for each cluster, we perform ordering and store the largest of these distances:
\begin{eqnarray}
 \mathcal{B}_{\max \min RU} = \max(\mathcal{B}_{z_{\max \min RU}})
\end{eqnarray}

Then, we select the cluster and the relay that are associated with
this distance to transmit simultaneously $M$ packets from the
particular buffer to the selected cluster. Considering imperfect
CSI, the estimated channel matrix $\mathbf{\hat{H}}$ is assumed,
instead of $\mathbf{H}$ in (\ref{eq:7}) and (\ref{eq:8a}). We remark
that other resource allocation techniques such as power allocation
and relays \cite{jpais,jlrpa,altpard,jicrs,jpba} can also be
considered.

\subsection{Comparison of metrics and choice of transmission mode}

After computing all the metrics associated with the UR and RU
channels and finding $\mathcal{B}_{\max \min UR}$ and
$\mathcal{B}_{\max \min RU}$, we compare these parameters and choose
the transmission mode:

    - If $\frac{\mathcal{B}_{\max \min UR}}{\mathcal{B}_{\max \min RU}} \geq C$, we select "MA mode",

 - Otherwise, we select "BC mode".\\

where $C =\frac{E[\mathcal{B}_{\max \min UR}]}{E[\mathcal{B}_{\max
\min RU}]}$. Thus, the probability of a relay being selected for
transmission is close to the probability of a relay being selected
for reception, and, consequently, the protocol works in a balanced
way, even for asymmetric channels.

\section{Analysis}
In this section, we first analyze the proposed MW-Max-Link in terms of PEP. Then, an approximated expression for the sum rate of the proposed protocol is derived and the complexity of the proposed and existing schemes are also presented.

\subsection{Pairwise Error Probability}

The equations for $\mathcal{B}_{UR_{n}}$, $\mathcal{B}_{R_{n}U_1}$
and $\mathcal{B}_{R_{n}U_2}$ may be simplified by making
$\mathcal{B}=  E_s/M \times \mathcal{B'}$,  where
$\mathcal{B'}=\norm{\mathbf{H}(\mathbf{x}_i-\mathbf{x}_j)}^2$ in
(\ref{eq:7}) and (\ref{eq:8a}). The PEP considers the error event
when $\mathbf{x}_i$  is transmitted and the detector computes an
incorrect $\mathbf{x}_j$  (where "$i$" is different from "$j$"),
based on the received symbol \cite{f41}. The PEP is given by
\begin{eqnarray}
\begin{split}
\mathbf{P}(\mathbf{x}_i \rightarrow \mathbf{x}_j | \mathbf{H})&= Q\left(\sqrt{\frac{E}{2 N_0M} \mathcal{B'}}\right)
  \label{eq:19}
\end{split}
\end{eqnarray}
where $N_0$ is the power spectrum density of the AWGN. The PEP will have its maximum value for the minimum value of $\mathcal{B'}$ (PEP worst case).
Thus, for cooperative transmissions, an approximated expression for computing the PEP worst case ($\mathcal{B'}_{\min}$) in each time slot (regardless of whether it is an UR or RU channel) is given by
\begin{eqnarray}
\mathbf{P}(\mathbf{x}_i \rightarrow \mathbf{x}_j | \mathbf{H})\approx 1- \left(1-Q\left(\sqrt{\frac{E}{2 N_0M} \mathcal{B'}_{\min}}\right)\right)^2
  \label{eq:102}
\end{eqnarray}

The extended MMD relay selection algorithm maximizes the metric
$\mathcal{B'}_{\min}$ and, consequently, minimizes the PEP worst
case in the proposed MW-Max-Link scheme.

\subsection{Sum Rate}

The sum rate of a given system is upper bounded by the system
capacity. In the MW-Max-Link scheme, as $R_g$ may be different from
$R_f$, its capacity is given by \cite{f1, f15}:
\begin{eqnarray}
  C_{DF}=\frac{1}{2} \min \{ I_{DF}^{UR_g},I_{DF}^{R_fU}\},
  \label{eq:150}
\end{eqnarray}
where the first and second terms in (\ref{eq:150}) represent the
maximum rate at which $R_g$ can reliably decode the messages
transmitted by the selected cluster ($U_1$ and $U_2$) and  at which
the selected cluster can reliably decode the estimated messages
transmitted by $R_f$, respectively.

For the mutual information between $U_1$ and $U_2$ and $R_g$, considering perfect CSI, we have
\begin{eqnarray}
\begin{split}
 I_{DF}^{UR}&=I_{DF}(\mathbf{x};\mathbf{y}_{u,r} | \mathbf{H}_{u,r}),\\
&= E[ \log_2 \det(\mathbf{I}+
\mathbf{H}_{u,r}\mathbf{Q}_{u,r}\mathbf{H}_{u,r}^H/N_0)],
  \label{eq:26}
\end{split}
\end{eqnarray}
where $\mathbf{H}_{u,r}$ represents a $2M\times 2M$ channel matrix and
\begin{eqnarray}
\begin{split}
E[\mathbf{y}_{u,r} \mathbf{y}_{u,r}^H]&=E[( \mathbf{H}_{u,r}\mathbf{x}+\mathbf{n}_{r})( \mathbf{H}_{u,r}\mathbf{x}+\mathbf{n}_{r})^H],\\
&= E[ \mathbf{H}_{u,r}\mathbf{x}(\mathbf{x})^H(\mathbf{H}_{u,r})^H+\mathbf{n}_{r}(\mathbf{n}_{r})^H],\\
&= \mathbf{H}_{u,r} \mathbf{Q}_{u,r}(\mathbf{H}_{u,r})^H+N_0 \mathbf{I}
  \label{eq:28}
\end{split}
\end{eqnarray}
where $\mathbf{Q}_{u,r}=
E[\mathbf{x}(\mathbf{x})^H]=\mathbf{I}~\frac{ E}{M}$ is the
covariance matrix of the transmitted symbols. Note that the vectors
$\mathbf{x}$ are formed by independent and identically distributed
(i.i.d.) symbols. The same reasoning can be applied to
$I_{DF}^{RU}$:
\begin{eqnarray}
I_{DF}^{RU}= E[ \log_2 \det(\mathbf{I}+
\mathbf{H}_{r,u}\mathbf{Q}_{r,u}\mathbf{H}_{r,u}^H/N_0)]
  \label{eq:33}
\end{eqnarray}
where $\mathbf{Q}_{r,u}=\mathbf{I}~ \frac{E}{M}$ and
$\mathbf{H}_{r,u}$ represents an $M\times M$ channel matrix.

To compute the sum rate of the MW-Max-Link scheme, instead of
(\ref{eq:150}),  we consider an approximated expression for the sum
rate in each time slot, depending on the kind of transmission. Then,
in the case of a time slot $i$ selected for UR transmission, the
approximated sum rate is given by
\begin{eqnarray}
R_i^{UR}\approx \frac{1}{2}~  \log_2 \det(\mathbf{H}_{u,r}
\mathbf{Q}_{u,r}\mathbf{H}_{u,r}^H/N_0 +\mathbf{I})
  \label{eq:35}
\end{eqnarray}

Moreover, in the case of a time slot $i$ selected for RU transmission, the approximated sum rate is given by
\begin{eqnarray}
R_i^{RU_{1(2)}}\approx \frac{1}{2}~  \log_2
\det(\mathbf{H}_{r,u_{1(2)}}
\mathbf{Q}_{r,u_{1(2)}}\mathbf{H}_{r,u_{1(2)}}^H/N_0 + \mathbf{I})
  \label{eq:36}
\end{eqnarray}

So, the average sum rate ($R$)  of the MW-Max-Link scheme may be approximated by
\begin{eqnarray}
R\approx \frac{\sum_{i=1}^{n_{UR}} R_i^{UR}+\sum_{i=1}^{n_{RU}} (R_i^{RU_1}+ R_i^{RU_2})}{n_{UR}+n_{RU}},
  \label{eq:38}
\end{eqnarray}
where $n_{UR}$ and $n_{RU}$ represent the number of time slots selected for UR and RU transmissions, respectively.

\subsection{Computational Complexity}

The complexity of the proposed MW-Max-Link, TW-Max-Link and the
existing TW-Max-Min scheme (here adapted for multi-antenna systems)
are associated with the complexity of the MMD protocol \cite{f41}.
The number $\mathcal{X}$  of calculations of the  metric
$\mathcal{B}$ for each channel matrix $ \mathbf{H}$ is given by
\begin{eqnarray}
\mathcal{X}=  \sum_{i=1}^{M'} 2^{i-1}  W^i C_i^{M'}
\label{eq:23}
\end{eqnarray}
where $M'=2M$  in the case of $\mathcal{B}_{UR_n}$, for the MA mode
($\mathcal{X}^{MA}$), and  $M'=M$ in the case of
$\mathcal{B}_{R_nU_1}$ and $\mathcal{B}_{R_nU_2}$, for the BC mode
($\mathcal{X}^{BC}$), $W$ is the number of different distances
between the constellation symbols. If we have BPSK, $W=1$, and QPSK,
$W=3$.
\begin{table}[!htb]
\centering
 \caption{Computational Complexity}
 \label{table2}
\begin{tabular}{l|ll}
\hline
Operations& MW-Max-Link& TW-Max-Min \cite{f40}\\
\hline
additions & $ZNM(2\mathcal{X^{BC}+X^{MA}}-3)$ & $\frac{NM}{2}(2\mathcal{X^{BC}+X^{MA}}-3)$\\
\hline
 multiplications &  $ZNM(2\mathcal{X^{BC}+X^{MA}})$ & $\frac{NM}{2}(2\mathcal{X^{BC}+X^{MA}})$ \\
\hline
\end{tabular}
\end{table}
Table \ref{table2} shows the complexity of the proposed MW-Max-Link and the existing TW-Max-Min, for $Z$ clusters, $N$ relays, $M$ antennas at the user nodes and $2M$ antennas at the relays. Note that TW-Max-Link is a special case of  MW-Max-Link, for a single two-way relay channel ($Z=1$). The complexity of MW-Max-Link is equal to the complexity of the adapted TW-Max-Min, multiplied by $2Z$.

\section{Simulation Results}

This section illustrates and discusses the
simulation results of the proposed MW-Max-Link, the TW-Max-Link and the adapted TW-Max-Min \cite{f40}, using the extended MMD relay selection criterion. The transmitted signals belong to BPSK
constellations. The use of high order constellations as QPSK and 16-QAM was not included in this work but can be considered elsewhere. We tested the performance for different $J$, but found that $J=6$ packets is sufficient to ensure a good performance in TW-Max-Link and  MW-Max-Link.
We also assume unit power channels ($\sigma_{ u,r}^2$ $=$ $\sigma_{ r,u}^2$ $= 1$) and $N_0 =1$. The transmit signal-to-noise ratio SNR ($E/N_0$) ranges
from 0 to 10 dB, where $E$ is the total energy transmitted by each user or the relay. The performances of the schemes
were tested for $10000 M$ packets, each containing 100 symbols. For imperfect CSI, the estimated channel matrix $\mathbf{\hat{H}}$ is assumed instead of $\mathbf{H}$: $\mathbf{\hat{H}}$=$\mathbf{H}$+$\mathbf{H}_e$, where the variance of the $\mathbf{H}_e$ coefficients is given by $\sigma_e^2=\beta E^{-\alpha}$ ($\beta \geq 0$ and $0 \leq \alpha \leq 1$) \cite{f16}.

\begin{figure}[!h]
\centering
\includegraphics[scale=0.53]{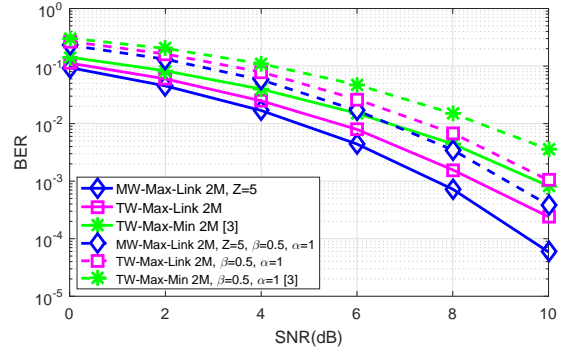}
\caption{ BER performance versus SNR.}
\label{fig:pepMaxlink}
\end{figure}

Fig. \ref{fig:pepMaxlink} shows the BER performance of the
MW-Max-Link for $Z = 5$, TW-Max-Link and TW-Max-Min protocols, for
$M = 2$, $N = 10$, BPSK, perfect and  imperfect CSI  ($\beta=0.5$
and $\alpha=1$). The performances of the MW-Max-Link are
considerably better than those of TW-Max-Link and TW--Max-Min for
the total range of SNR values tested, both for perfect and imperfect
CSI.

\begin{figure}[!h]
\centering
\includegraphics[scale=0.53]{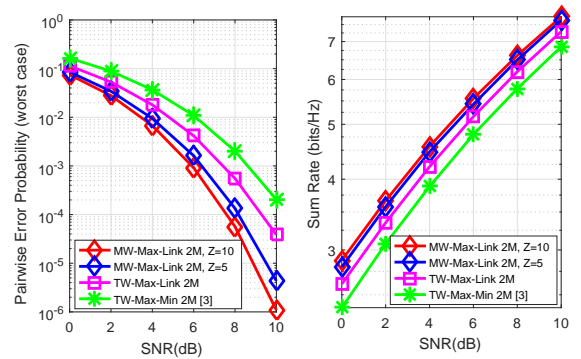}
\caption{PEP and Sum-Rate performances versus SNR.}
\label{fig:berMMDMaxLink}
\end{figure}

Fig. \ref{fig:berMMDMaxLink} shows the PEP and the Sum Rate
performances, for BPSK and Gaussian distributed signals,
respectively,  of the MW-Max-Link (for $Z=5$ and $Z=10$),
TW-Max-Link and TW-Max-Min  protocols, for $M = 2$, $N = 10$ and
perfect CSI. The performances of the MW-Max-Link are very close, and
considerably better than those of TW-Max-Link and TW--Max-Min  for
the total range of SNR values tested, as MW-Max-Link selects the
best links among $Z$ clusters and $N$ relays.

\section{Conclusions}

In this paper, we have presented a relay-selection strategy for multi-way cooperative
multi-antenna systems that is aided by a central processor node, where a cluster formed by two users is selected to  simultaneously transmit to each other with
the help of relays. In particular, the proposed multi-way relay selection strategy selects the best link, exploiting the use of  buffers and PLNC, that is called MW-Max-Link. The proposed MW-Max-Link was evaluated experimentally and outperformed the TW-Max-Link and the existing TW-Max-Min scheme. The use of a central processor node and buffers in the relays is presented as a promising relay selection technique and a framework for multi-way protocols.

\end{document}